%% file: main.tex
\documentclass{article}
\usepackage{spconf,graphicx}
\usepackage{amsmath}

\usepackage{subfigure}
\usepackage{xcolor}         
\usepackage{xcolor,colortbl}
\usepackage{CJKutf8}
\usepackage{amsfonts}
\usepackage{CJK}
\usepackage{tikz}
\usepackage{url}
\usepackage{multirow}
\usepackage{booktabs} 
\usepackage{tablefootnote}
\usepackage{graphicx}

\title{ERNIE-SAT: Speech and Text Joint Pretraining \\for Cross-Lingual Multi-Speaker Text-to-Speech}
\newcommand{\specificthanks}[1]{\@fnsymbol{#1}}

\usepackage[symbol]{footmisc}

\makeatletter
\def\@name{ \emph{Xiaoran Fan$^{1\dagger}$\thanks{~$^\dagger$Work done during internship at Baidu Inc.}, Chao Pang$^{\star}$, Tian Yuan$^{\star}$, He Bai$^{2\dagger}$, Renjie Zheng$^{\star}$, Pengfei Zhu$^{\star}$, Shuohuan Wang$^{\star}$}
\\
\emph{Junkun Chen$^{3\dagger}$, Zeyu Chen$^{\star}$, Liang Huang$^{3}$, 
Yu Sun$^{\star}$, Hua Wu$^{\star}$}
}
\makeatother
\address{
$^{\star}$Baidu Inc., Beijing, China \\
$^{1}$University of Chinese Academy of Sciences, Beijing, China\\
$^{2}$University of Waterloo, Waterloo, ON, Canada\\
$^{3}$Oregon State University, Corvallis, OR, USA}

\begin{document}
\maketitle
\begin{abstract}
Speech representation learning has improved both speech understanding and speech synthesis tasks for single language.
However, its ability in cross-lingual scenarios has not been explored.
In this paper, we extend the pretraining method for cross-lingual multi-speaker speech synthesis tasks, including cross-lingual multi-speaker voice cloning and cross-lingual multi-speaker speech editing.
We propose a speech-text joint pretraining framework, where we randomly mask the spectrogram and the phonemes given a speech example and its transcription. By learning to reconstruct the masked parts of the input in different languages, our model shows great improvements over speaker-embedding-based multi-speaker TTS methods.
Moreover, our framework is end-to-end for both the training and the inference without any finetuning effort.
In cross-lingual multi-speaker voice cloning and cross-lingual multi-speaker speech editing tasks, our experiments show that our model outperforms speaker-embedding-based multi-speaker TTS methods.\footnote{Audio demos can be found in: \url{https://www.notion.so/ERNIE-SAT-DEMO-6f3ef7fbea944d9db46ba2770ab6693d}

\quad Code and models are available at: \url{https://github.com/PaddlePaddle/ERNIE}}

\end{abstract}
\begin{keywords}
cross-lingual speech synthesis, multi-speaker text-to-speech, speech pretraining
\end{keywords}

\input{intro.tex}

\input{model.tex}

\input{exp.tex}

\input{conclusion.tex}

\bibliographystyle{IEEEbib}
\bibliography{main}

\end{document}

%% file: intro.tex
\vspace{-0.3cm}
\section{Introduction}
\vspace{-0.2cm}
\label{sec:intro}

Speech synthesis has made significant progress~\cite{wang2017tacotron,sotelo2017char2wav,ren2019fastspeech,2021FastSpeech} in recent years.
But it is still challenging to synthesize speech in different languages, a.k.a. cross-lingual TTS. 

Cross-lingual TTS aims to build a system that is capable of synthesizing speech in a specific language that is not spoken by the target speaker. 
This technology has a wide range of application scenarios that can benefit various fields, such as speech translation and simultaneous interpretation systems. 
However, there is no existing training data for such a task, which is expensive and difficult to collect.

A highly effective and intuitive baseline is to build a pipeline system~\cite{chen2019cross,cai2020cross,jia2018transfer}: extracting the speaker feature with an external speaker encoder model first, then synthesizing the target language speech with a target language multi-speaker TTS model. 
Ideally, the speaker encoder model should be trained with the multi-speaker speech data in both the source and the target language. 
The TTS model should be trained with both the multi-speaker speech-text data in the target language and the target language speaker embeddings.
But collecting a large amount of multi-speaker, multi-lingual data to train the speaker encoder is very difficult for low-resource languages.
Most systems just use the target language speaker model to extract the source language speakers' embeddings.
In this case, the cross-lingual TTS problem is treated as a new speaker TTS problem, while the source language speaker's voice and the global rhythm of the source language may not be captured by the target language speaker model.
Moreover, this is not an end-to-end system in terms of training and inference.

\vspace{-1.5pt}
On the other hand, pretraining models have shown great effectiveness in speech-processing tasks and are promising for low-resource scenarios.
But most of the recent pretraining work mainly focuses on speech representation learning instead of speech synthesis.
For example, Wav2vec 2.0~\cite{baevski2020wav2vec}, Hubert~\cite{hsu2021hubert}, and SLAM~\cite{bapna2021slam} have the optimization goal of learning discrete units, which are not able to handle speech synthesis problems.
MAM~\cite{chen2020mam} and FAT-MLM~\cite{zheng2021fused} propose to do self-supervised learning by reconstructing the masked spectrogram, but the reconstructed spectrogram is far from the quality of speech synthesis.
Most recently, A$^3$T~\cite{bai20223} reconstructs the masked acoustic signals with high quality for TTS tasks. 
However, A$^3$T can only handle monolingual speech and does not support cross-lingual synthesis.

\vspace{-1.5pt}
In this paper, we propose a general cross-lingual
speech-text joint pretraining framework ERNIE-SAT, which learns to reconstruct the masked speech and text jointly during the pretraining.
And We propose a non-overlapping masking strategy for these two modalities.
By pretraining with the monolingual data in English and Chinese, our models can be adopted as both the cross-lingual multi-speaker voice cloning system and the cross-lingual multi-speaker speech editing system without finetuning. Our results show that our model can generate high-quality speech for new speakers without any external speaker encoding model and even outperform the pipeline system equipped with speaker embeddings.

We make the following contributions: 

\begin{itemize}
\vspace{-0.1cm}
    \vspace{-5pt}
    \item We propose a non-overlapping masking strategy of pretraining based on forced alignment preprocessing of speech and text, which can reconstruct masked spectrograms with higher quality.
    \vspace{-7pt}
    \item ERNIE-SAT is able to perform cross-lingual speech synthesis for new speakers and outperforms speaker-embedding-based multi-speaker TTS methods.
    \vspace{-7pt}
    \item ERNIE-SAT is ready to use without finetuning once pretrained, this is the first time that only a single model (ERNIE-SAT) can be applied to voice cloning and speech editing in cross-lingual multi-speaker scenario.
    \vspace{-7pt}
\end{itemize}

%% file: model.tex
\vspace{-0.5cm}
\section{Method}
\label{sec:model}

\vspace{-0.4cm}
\subsection{ERNIE-SAT}
\vspace{-0.1cm}

\textbf{Raw Input Features} ERNIE-SAT takes speech and transcription tuples as input,
denotes as 
$D_{\mathrm{s}, \mathrm{x}} = \{\langle \mathrm{s}, \mathrm{x} \rangle \}$,  
where $\mathrm{s}=(s_1, ... , s_{|s|})$ is a sequence of acoustic features~(spectrogram or mel-spectrogram of the speech), and each $s_i \in \mathbb{R}^{d_{\text s}}$ represents the frame-level speech feature. $\mathrm{x}=(x_1, ... , x_{|x|})$ is the sequence of the corresponding transcription. 

\begin{figure}[t!]
    \centering
    \begin{tabular}{c}
    \vspace{-0.5cm}
    \\
    \begin{minipage}[b]{0.75 \linewidth}
    \begin{center}
    \includegraphics[width=7.cm]{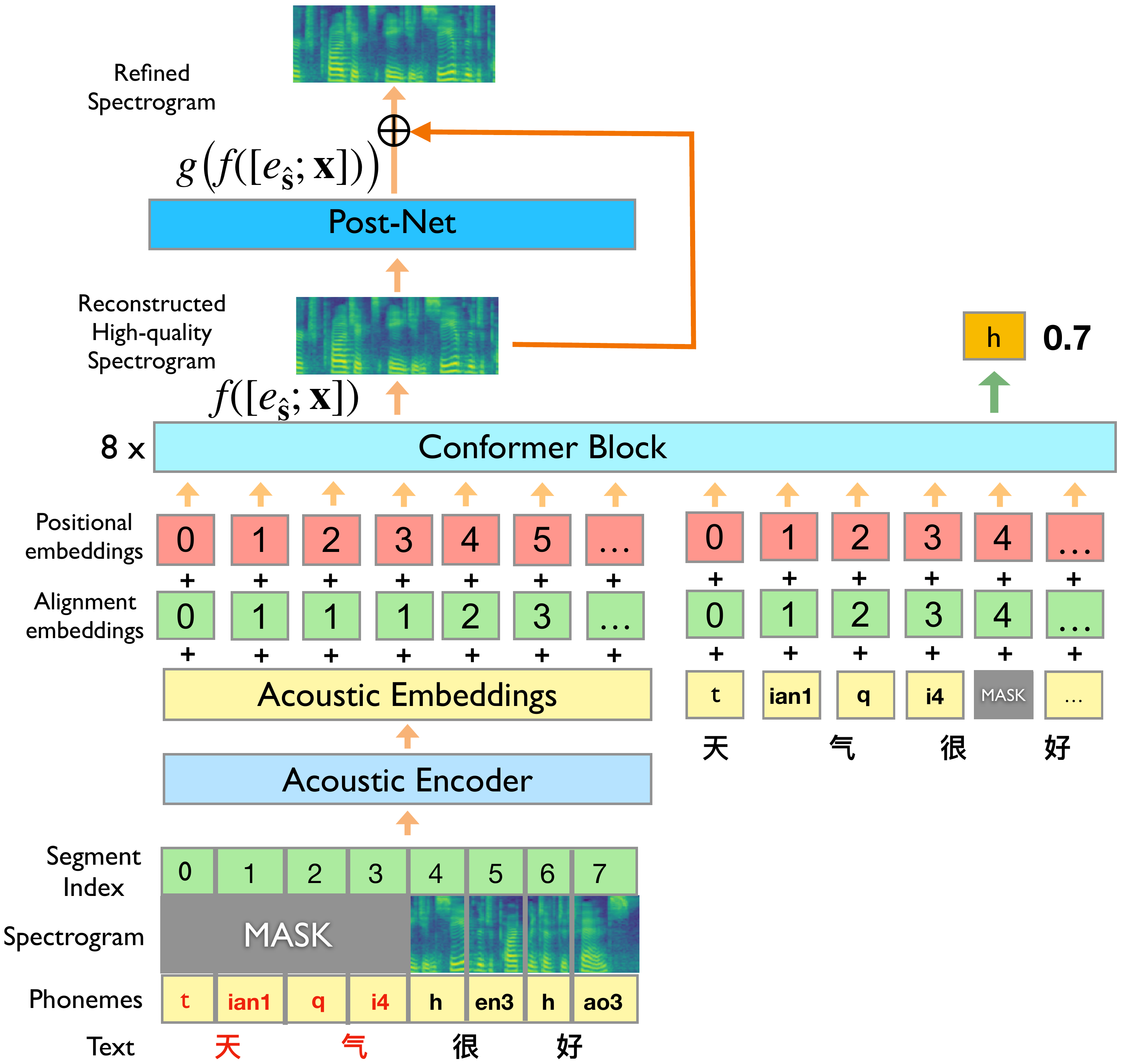}
    \end{center}
    \end{minipage}
    \end{tabular}
    \caption{ERNIE-SAT model architecture. }
    \vspace{-0.3cm}
    \label{fig:speech_edit}
\end{figure}

\noindent \textbf{Masking Strategy} As shown in Fig.~\ref{fig:speech_edit}, we first use a forced aligner to pre-process the speech and transcription to get the alignment information. 
Then, we propose a non-overlapping masking strategy for both the speech and the text. 
We mask several random spans on $\mathrm{s}$ with the masking function $\hat{\mathrm{s}} \sim  \text{Mask}_{\text{span}}(\mathrm{s}, \lambda)$. 
This function replaces several random spans of $\mathrm{s}$ with the same number of a random initialized masking vector $\epsilon_{\mathrm{s}} \in \mathbb{R}^{d_{\text s} }$. 
$\lambda$ controls the portion of the phonemes whose corresponding frames to be masked.
We follow the random span selection from T5~\cite{raffel2020exploring} and the masking portion $\lambda=80\%$ from $\text{A}^3\text{T}$~\cite{bai20223}.
Similarly, for text masking, based on the alignment information, we randomly mask half of the phonemes that are not selected above with a random masking function $\hat{\mathrm{x}} \sim \text{Mask}_{\text{phoneme}}(\mathrm{x}, (1-\lambda)/2)$, where $\text{Mask}_{\text{phoneme}}(\cdot)$.
The masked phonemes are replaced with another random initialized vector $\epsilon_{\mathrm{x}} \in \mathbb{R}^{d_{\text x} }$.

\noindent \textbf{Input Layer} After masking, the masked spectrogram frames are replaced into $\epsilon_{\mathrm{s}}$. 
Then we use a nonlinear feed-forward layer as the acoustic encoder, to encode $\hat{\mathrm{s}}$ into the acoustic embeddings  $\mathrm{e}_{\hat{\mathrm{s}}}$.
For the text, unmasked phonemes are encoded into phoneme embeddings while the masked phonemes are replaced with $\epsilon_{\mathrm{x}}$.
In addition to the vanilla position encoding from Transformer, we incorporate the alignment embedding~\cite{bai20223, bai2021segatron} to strengthen the interaction between the speech and text.
Before the interaction between speech and text, the vector of speech is represented as $\mathrm{e}_{s_i}+\mathrm{e}_{\text{pos}_i}+\mathrm{e}_{\text{aln}_i}$ and the vector of text is represented as $\mathrm{e}_{x_i}+\mathrm{e}_{\text{pos}_i}+\mathrm{e}_{\text{aln}_i}$.

\noindent \textbf{Encoder} We adopt Conformer~\cite{gulati2020conformer} as the backbone of our encoder. 
The acoustic and text embeddings are concatenated together as the input of our Encoder block.

\vspace{0.05cm}
\noindent \textbf{Training Loss} ERNIE-SAT is trained by reconstructing the masked spectrogram and phonemes. 

The acoustic reconstruction loss is:
\vspace{-0.15cm}
\begin{equation}
\begin{aligned}
\ell_{\mathrm{s}}(D_{\mathrm{s},  \mathrm{x}}) = \displaystyle\sum_{\langle\mathrm{s},  \mathrm{x}\rangle \in D_{\mathrm{s},  \mathrm{x}}}  &\| f ([e_{\hat{\mathrm{s}}}; \mathrm{x}]) + g \big( f ([e_{\hat{\mathrm{s}}}; \mathrm{x}])\big) - \mathrm{s}  \|_1
\\
&+\|  f ([e_{\hat{\mathrm{s}}}; \mathrm{x}])  - \mathrm{s}  \|_1
\label{eq:reconstruct}
\end{aligned}
\end{equation}
where $g$ is a Post-Net~\cite{wang+:2017} to recover the ground-truth
signal from the encoder's output $f([e_{\hat{\mathrm{s}}}; \hat{\mathrm{x}}])$.
We use mean absolute error~(MAE) to measure the distance
between $s$ and the reconstructed spectrogram.

We use the cross-entropy loss for text reconstruction:
\vspace{-0.2cm}
\begin{equation}
\begin{aligned}
\ell_{\mathrm{x}}(D_{\mathrm{s},  \mathrm{x}}) = - \displaystyle\sum_{\langle\mathrm{s},  \mathrm{x}\rangle \in D_{\mathrm{s},  \mathrm{x}}} log p(\mathrm{x} | [e_{\hat{\mathrm{s}}}; \mathrm{x}])
\end{aligned}
\end{equation}

The final loss is the combination of the speech and text:
\vspace{-0.25cm}
\begin{equation}
\begin{aligned}
\ell_{\mathrm{ERNIE-SAT}}(D_{\mathrm{s},  \mathrm{x}}) = \ell_{\mathrm{s}}(D_{\mathrm{s},  \mathrm{x}}) + \ell_{\mathrm{x}}(D_{\mathrm{s},  \mathrm{x}}) 
\end{aligned}
\end{equation}

\vspace{-0.1cm}
\noindent \textbf{Cross-lingual pretraining} We mix Chinese and English phonemes together and train on both Chinese and English monolingual datasets. During the training process, we randomly sampled data from the Chinese and English monolingual datasets to form batches and randomly alternated training, based on the above model structure, the non-overlapping masking strategy, and the training objective.

\vspace{-0.3cm}
\subsection{Cross-Lingual Multi-Speaker Voice Cloning}
\vspace{-0.1cm}
For cross-lingual voice cloning, we use the prompt-based decoding method~\cite{bai20223}. The specific implementation is as follows:
we first concatenate the source language prompt text and target language text together, then we use a duration predictor to get the length of the target speech and append the same amount of \texttt{[MASK]} vectors after the prompt speech.
By feeding the speech and text inputs into our ERNIE-SAT model, the masked spectrogram is reconstructed, and then the target language audio is obtained.

\vspace{-0.3cm}
\subsection{Cross-Lingual Multi-Speaker Speech Editing}
\vspace{-0.1cm}
Inspired by~\cite{tan2021editspeech} and A$^3$T in the monolingual speech editing scenario, we propose a new task to probe the cross-lingual synthesis effect of our model, which is cross-lingual multi-speaker speech editing. Moreover, cross-lingual speech editing can support making the content conveyed by the speaker easier to understand in certain speech translation scenarios without degrading the quality and naturalness of the edited speech.
This task is to synthesize speech in the target language by editing some phrases in the source language text into the target language through a synthesizer, where the editing operations include insertion, deletion, and replacement.

The implementation can be summarized in the following steps: (1) Locate the edited text 
$\mathrm{x}$ and using an additional duration predictor to predict the duration of the edited phonemes ${\mathrm{d}}$
(2) Then, based on the correspondence between speech and text, locate the segment to the spectrogram. Using length of $\sum_{i=1}^{n} \hat{\mathrm{d}}_i \cdot \mathtt{sr} / \mathtt{h}$\footnote{$\mathtt{sr}$ stands for sample rate and $\mathtt{h}$ stands for hop size.} the \texttt{[MASK]} frames inserted into the unmodified spectrogram context. (3) By inputting the concatenated edited spectrogram with the masked part and the edited text, ERNIE-SAT will reconstruct the spectrogram $f_{\hat{\mathrm{s}}}$ of these masked frames. (4) Use the vocoder to generate the waveform of this spectrogram, and replace the original speech with $f_{\hat{\mathrm{s}}}$ as the final edited speech.

\begin{figure}[t!]
    \centering
    \begin{tabular}{c}
    \\
    \begin{minipage}[b]{.7 \linewidth}
    \begin{center}
    \includegraphics[width=6.5 cm]{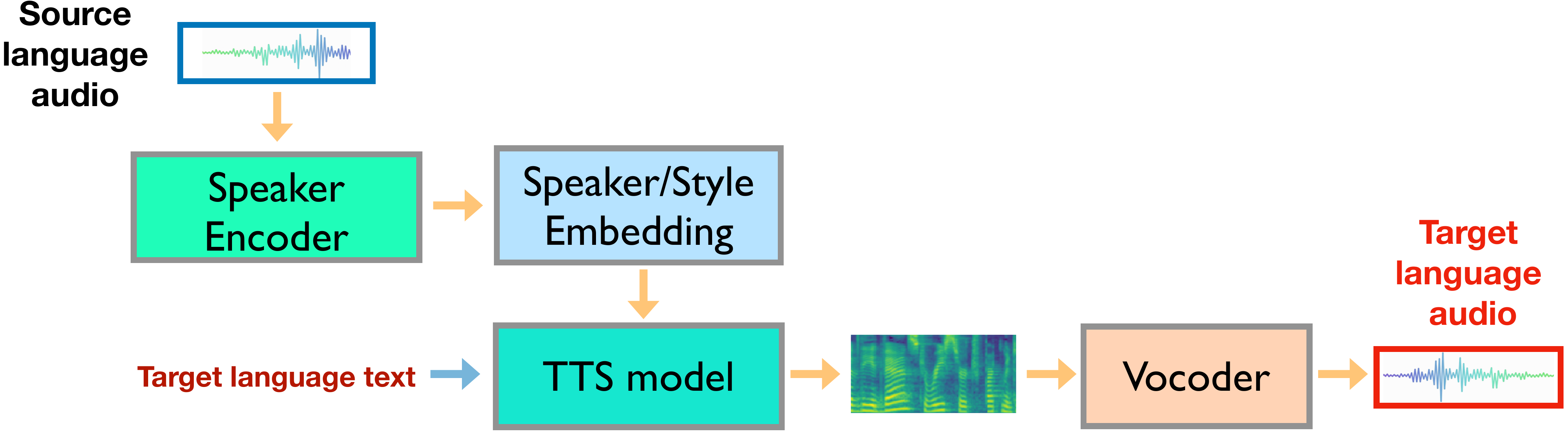}
    \end{center}
    \end{minipage}
    \end{tabular}
    \caption{Speaker/Style embedding-based method.}
    \vspace{-0.5cm}
    \label{fig: one-shot-tts}
\end{figure}

\vspace{-0.15cm}

%% file: exp.tex
\vspace{-0.3cm}
\section{Experiments}
\label{sec:experiment}

\input{ablation_plot.tex}

\begin{figure}[t!]
    \centering
    \begin{tabular}{c}
    \\
    \begin{minipage}[b]{.7 \linewidth}
    \begin{center}
    \includegraphics[width=4.2cm]{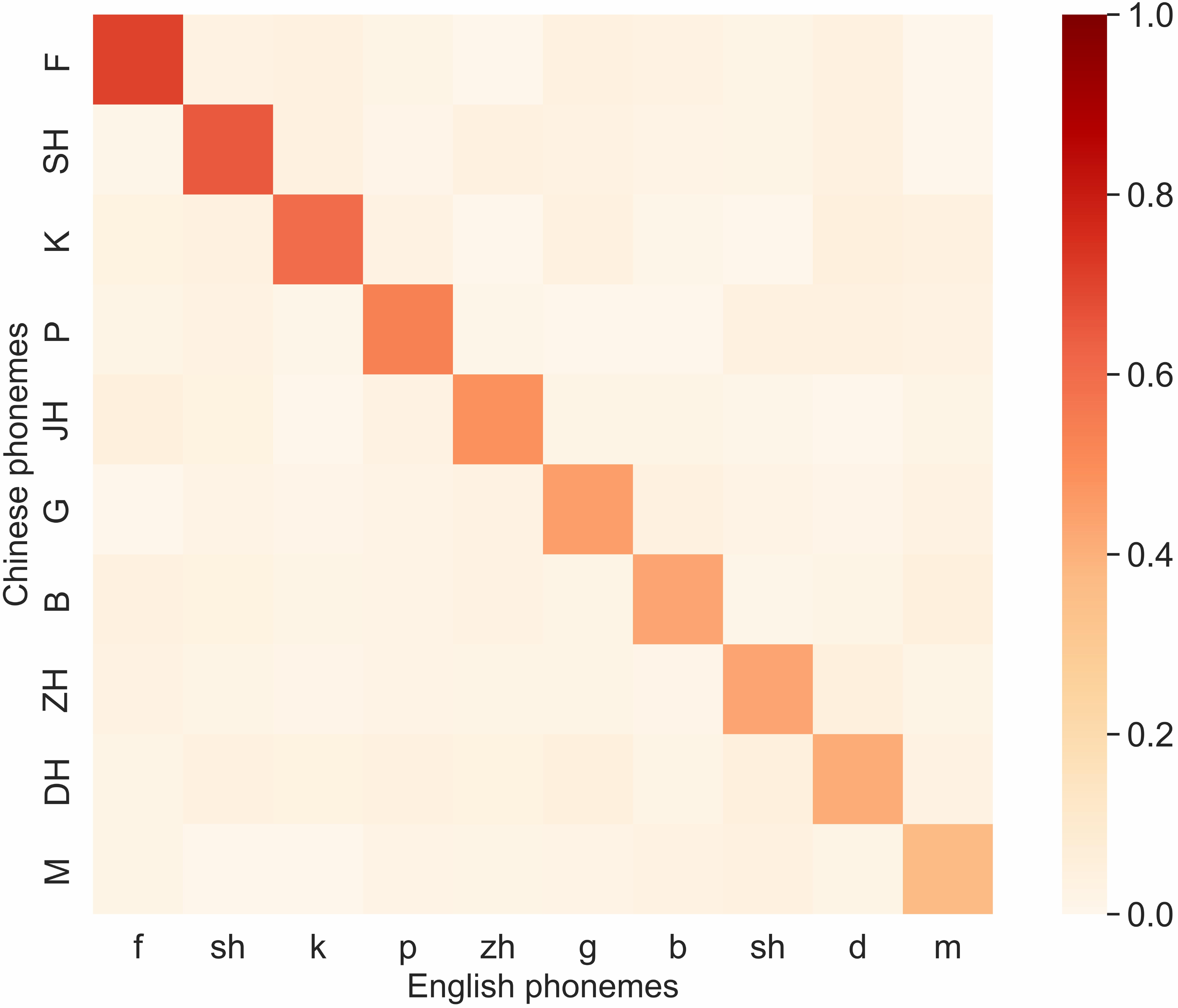}
    \end{center}
    \end{minipage}
    \end{tabular}
    \caption{Context-aware Phoneme Mappings (sampled).}
    \vspace{-0.3cm}
    \label{fig: mapping}
\end{figure}

\vspace{-0.2cm}
\subsection{Datasets}
\vspace{-0.15cm}
In our experiments, We use two public datasets to train the cross-lingual multi-speaker speech synthesis networks, AISHELL3 (Chinese multi-speaker TTS dataset) and VCTK (English multi-speaker TTS dataset). VCTK contains 44 hours of clean speech from 109 speakers. AISHELL3 contains 85 hours of high-quality speech and text data from 218 speakers from different accent regions in China. We downsampled all audios to 24 kHz.

\vspace{-0.3cm}
\subsection{Training Detail}
\vspace{-0.1cm}
In the speech preprocessing, we use a window size of 50 ms and a hop length of 12.5 ms to process all raw audio files and then extract the 80-dimensional log-Mel filter set using the Hann window function and STFT. The forced alignment and G2P are both carried out by HTK. There are 193 phonemes in Chinese and 73 phonemes in English. After de-duplication and merging, 262 cross-lingual phonemes are obtained.

Our model architecture: nonlinear feed-forward layer as the acoustic encoder, 8 layers of Conformer encoder, 5 layers Conv1d Post-Net. The convolution kernel sizes of the first 4-layer encoder and the next 4-layer encoder are 7 and 31, respectively.
For pretraining, we use the Adam optimizer with an initial learning rate of 1.0, warmup step of 4000, and a Noam learning rate scheduler. The batch size is set by batch-bin, which is dynamically adjusted during the training process. For cross-lingual speech synthesis systems, we use the publicly available duration predictor in FastSpeech 2 implemented in ESPnet and the Parallel-WaveGAN~\cite{yamamoto2020parallel} vocoder.

\vspace{-0.4cm}
\subsection{Baseline models}
\vspace{-0.1cm}
To the best of our knowledge, SV2TTS~\cite{jia2018transfer} is the best cross-lingual speech synthesis framework,  shown in Fig.~\ref{fig: one-shot-tts}, which is composed of three independently trained networks: (1) speaker encoder network; (2) sequence-to-sequence synthesis network; (3) vocoder network. For cross-lingual bi-directional synthesis, the above baseline systems require both the source language and the target language speaker verification models and synthesizers. Moreover, speaker verification trained on the speaker verification task using an independent dataset of pure speech from tens of thousands of speakers.

\vspace{-0.4cm}
\subsection{Cross-lingual Multi-Speaker Voice Cloning}
\vspace{-0.1cm}

For the cross-lingual multi-speaker synthesis, we adopt two baseline systems based on the SV2TTS framework.

To evaluate the ability of cross-lingual synthesizing, we randomly sampled unseen speakers in the VCTK and AISHELL3 test sets, containing a total of 20 utterances of unseen speakers. Each audio sample is listened by 10 subjects whose first language is Chinese and are well-educated in English. 
The subjects are asked to evaluate the quality and similarity of synthesized audio.
The results are shown in Tab.~\ref{table: cross_lingual_TTS_quality} and Tab.~\ref{table: cross_lingual_TTS_sim}. From these tables, we can see that even without speaker embedding (X-Vector)~\cite{snyder2018x}, Our ERNIE-SAT model outperforms other baseline systems in terms of speaker similarity and speech quality.

\begin{table}[ht!]
\vspace{-0.3cm}
	\centering
\resizebox{0.65\columnwidth}{!}{
		\begin{tabular}{l|c}
\hline
Model  & Unseen \\ \hline
 Tacotron 2 + X-vectors + GST & 3.33 $\pm$ 0.16   \\ \hline
FastSpeech 2 + X-vectors + GST   & 3.49 $\pm$ 0.14    \\ \hline
ERNIE-SAT              & 3.58 $\pm$ 0.14   \\ \hline
\end{tabular}
\vspace{-0.3cm}
}

\caption{ The MOS for \textbf{speech quality} on cross-lingual multi-speaker TTS with 95\% confidence intervals. }\label{table: cross_lingual_TTS_quality}
\end{table}

\vspace{-0.5cm}
\begin{table}[ht!]
	\centering
\resizebox{0.65\columnwidth}{!}{
		\begin{tabular}{l|c}
\hline
Model  & Unseen \\ \hline
 Tacotron 2 + X-vectors + GST    & 3.30 $\pm$ 0.17   \\ \hline
FastSpeech 2 + X-vectors + GST   & 3.45 $\pm$ 0.16   \\ \hline
ERNIE-SAT               & 3.53 $\pm$ 0.11   \\ \hline
\end{tabular}

}
\caption{ The MOS for \textbf{speaker similarity} on cross-lingual multi-speaker TTS with 95\% confidence intervals. }\label{table: 
cross_lingual_TTS_sim}
\end{table}

\vspace{-0.3cm}
\begin{table}[ht!]
	\centering
\resizebox{0.65\columnwidth}{!}{
		\begin{tabular}{l|c}
\hline
Model  & Unseen \\ \hline
 Tacotron 2 + X-vectors + GST    & 3.07 $\pm$ 0.18   \\ \hline
FastSpeech 2 + X-vectors + GST   & 3.37 $\pm$ 0.15   \\ \hline
ERNIE-SAT               & 3.45 $\pm$ 0.17   \\ \hline
\end{tabular}
\vspace{-0.1cm}
}

\caption{ The MOS for \textbf{speech quality} on cross-lingual multi-speaker speech editing with 95\% confidence intervals. }\label{table: cross_lingual_edit_quality}
\end{table}

\begin{table}[ht!]
	\centering
\resizebox{0.65\columnwidth}{!}{
		\begin{tabular}{l|c}
\hline
Model  & Unseen \\ \hline
 Tacotron 2 + X-vectors + GST    & 3.10 $\pm$ 0.19   \\ \hline
FastSpeech 2 + X-vectors + GST   & 3.34 $\pm$ 0.15   \\ \hline
ERNIE-SAT               & 3.51 $\pm$ 0.11   \\ \hline
\end{tabular}
\vspace{-0.1cm}
}
\caption{ The MOS for \textbf{speaker similarity} on cross-lingual multi-speaker speech editing with 95\% confidence intervals. }\label{table: cross_lingual_edit_sim}
\vspace{-0.3cm}
\end{table}

\vspace{-0.3cm}
\subsection{Cross-lingual Multi-Speaker Speech Editing}
\vspace{-0.2cm}
 We adopted the same baselines as the cross-lingual multi-speaker voice cloning. And the additional steps are using TTS models to generate the modified region and insert the generated waveform into the original waveform using a forced aligner.
We test cross-lingual multi-speaker speech editing task with VCTK dataset. Specifically, we use randomly 10 utterances of 10 speakers from ~\cite{tan2021editspeech} for cross-lingual speech editing.
The MOS scores in Tab.~\ref{table: cross_lingual_edit_quality} and Tab.~\ref{table: cross_lingual_edit_sim} indicate that our ERNIE-SAT model achieves better speech quality and similarity of the edited speech compared to the speaker-embedding-based multi-speaker TTS models.

\vspace{-0.3cm}
\subsection{Context-aware Cross-lingual Phoneme Mappings}
\vspace{-0.2cm}

Our model can generate Context-aware Cross-lingual Phoneme Mappings, because it is trained on both English and Chinese TTS datasets with mixed English and Chinese phonemes. For clarity, we only visualized the mapping between 10 Chinese and English phonemes in Fig.~\ref{fig: mapping}. From our pretrained model, we can calculate the pronunciation similarity of English phonemes and Chinese phonemes. For each English phoneme, we can find a corresponding similar Chinese phoneme, and the brighter the color, the higher similarity of these two phonemes.

\vspace{-0.3cm}
\subsection{Ablation Study}
\vspace{-0.2cm}

We conduct an ablation study for our pretraining task: spectrogram reconstruction. This task requires our models to predict the masked frames.
An example of different models’ reconstruction is shown in Fig.~\ref{fig:ablation}. The difference between our models is the masking strategy in the pretraining phase. By comparing Fig.~\ref{fig:ablation}(b) and Fig.~\ref{fig:ablation}(c), we can see the model that masks both speech and text in the pretraining phase reconstruct the spectrogram with clearer texture features and is closer to ground-truth, compared to masking only speech. 
This observation demonstrates the effectiveness of our proposed joint speech-text pretraining framework. This observation also proves that the non-overlapping mask strategy we proposed successfully enhanced the interaction between speech and text.

\vspace{-0.2cm}

%% file: ablation_plot.tex
\begin{figure*}[ht!]
\centering
\begin{tabular}{c}

\begin{minipage}[t]{.32 \linewidth}
\begin{center}
\vspace{-.5cm}
\subfigure[
Origin spectrogram from VCTK. 
]{
    \makebox[0.9 \linewidth][c]{
\includegraphics[height=3.8cm]{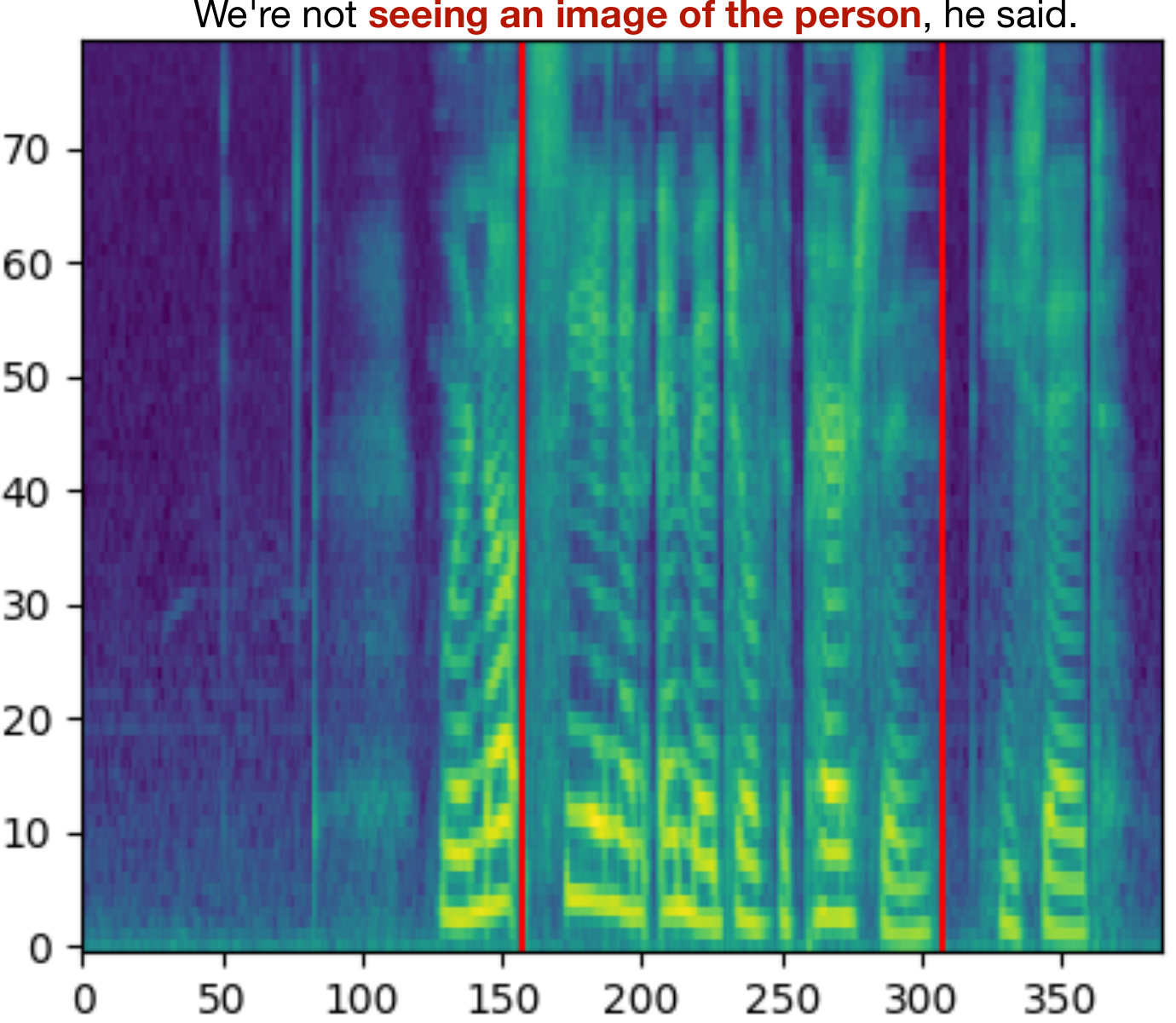}
\label{fig:ljs_ab_gt}
    }
}
\end{center}
\end{minipage}
\begin{minipage}[t]{.32 \linewidth}
    \begin{center}
\vspace{-.5cm}
    \subfigure[
    Reconstructed spectrogram based on ERNIE-SAT (\textbf{masking speech only} In the pretraining phase)
.
    ]{
    \makebox[0.9  \linewidth][c]{
    \includegraphics[height=3.8cm]{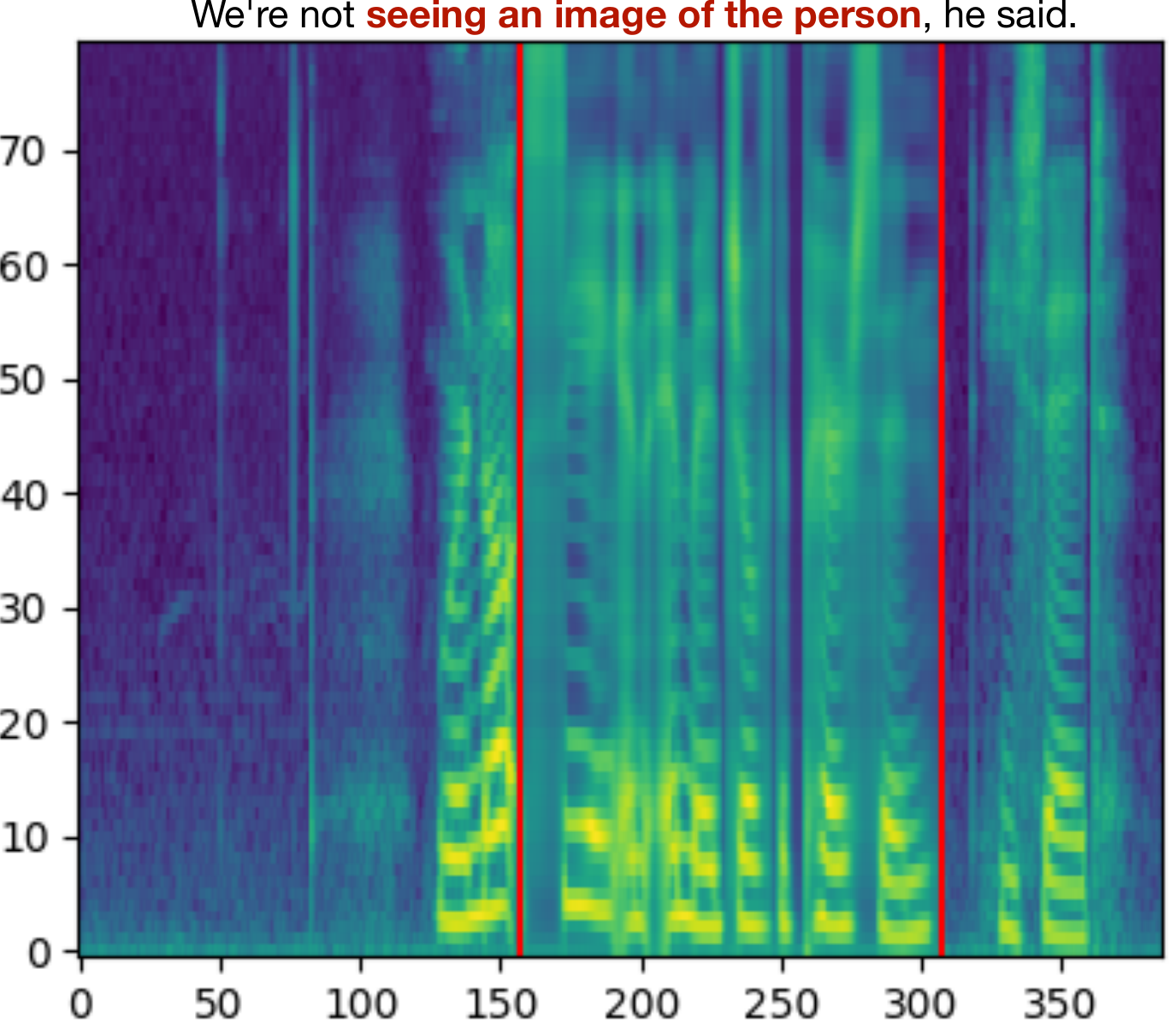}
    \label{fig:ljs_ab_b}
        }
    }
    \end{center}
\end{minipage}

\begin{minipage}[t]{.32 \linewidth}
    \begin{center}
\vspace{-.5cm}
    \subfigure[
   Reconstructed spectrogram based on ERNIE-SAT (\textbf{masking speech and text} In the pretraining phase).
    ]{
    \makebox[0.9 \linewidth][c]{
    \includegraphics[height=3.8cm]{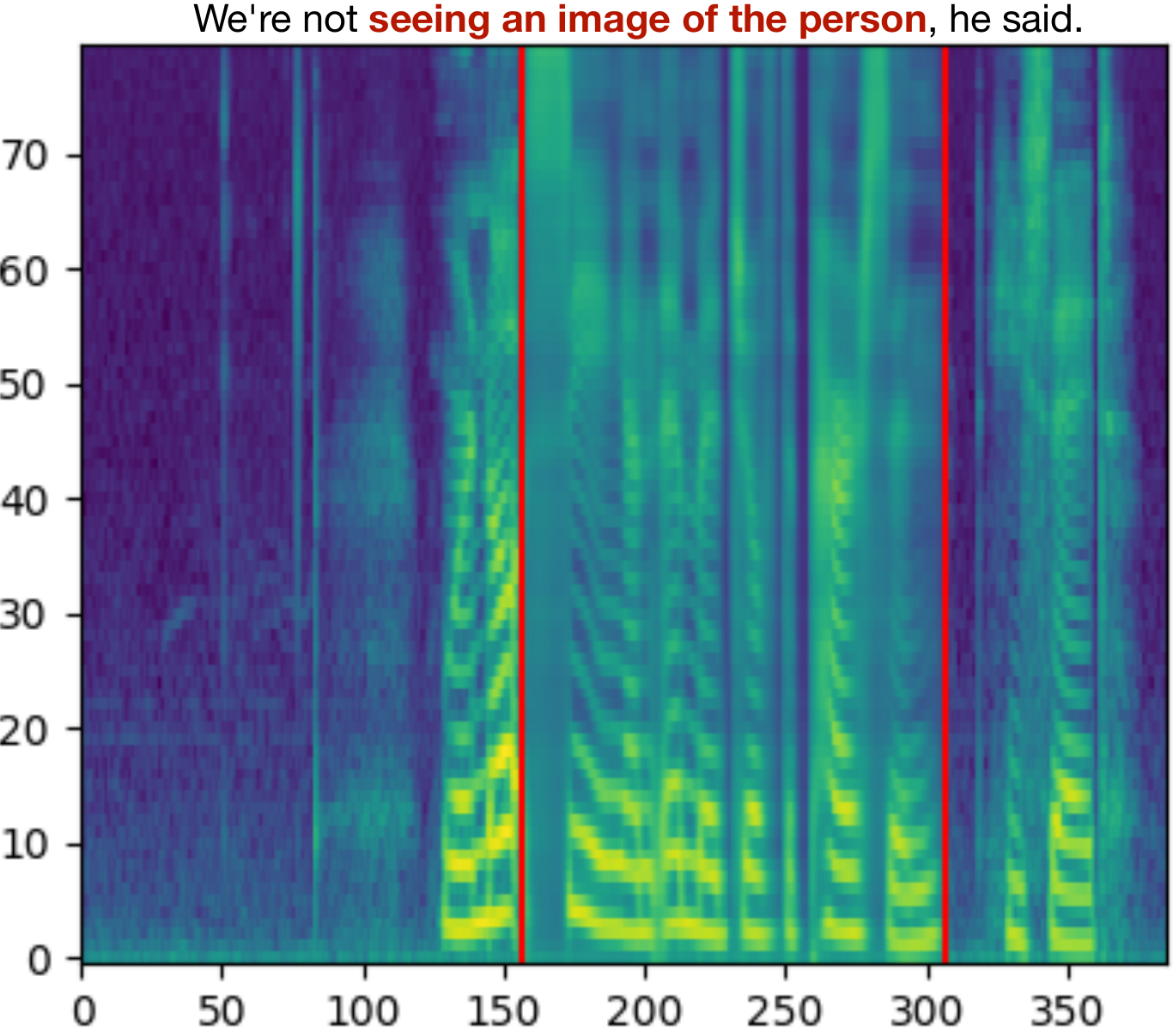}
    \label{fig:ljs_ab_b}
        }
    }
    \end{center}
\end{minipage}
\vspace{-0.3cm}
\end{tabular}
\caption{
    An example of ablation study for our pretraining task: spectrogram reconstruction. Original text is  
    ``We're not \textcolor{red}{seeing an image of the person} he said''. The portion with red box is 
    ``\textcolor{red}{seeing an image of the person}'' is masked in (b,c) subfigures.
}
\vspace{-0.3cm}
\label{fig:ablation}
\end{figure*}

%% file: conclusion.tex
\vspace{-0.3cm}
\section{Conclusion}
\vspace{-0.2cm}
In this paper, we propose a novel speech-text joint pretraining framework ERNIE-SAT, which is applicable to various types of cross-lingual multi-speaker speech synthesis tasks, such as cross-lingual multi-speaker voice cloning and cross-lingual multi-speaker speech editing.
Compared to the baseline models, our model is able to produce high-quality speech for seen-speakers and unseen-speakers even with a single model and without speaker embedding. 
Our results show that our proposed ERNIE-SAT model outperforms the speaker-embedding-based multi-speaker TTS models on unseen-speaker speech synthesis.